\documentclass[10pt,twocolumn,pra,aps,showpacs,nofootinbib,floatfix,nobibnotes]{revtex4-2}
\usepackage{amsmath, bm}
\usepackage{amssymb}
\usepackage{graphicx}
\usepackage{verbatim}
\usepackage[colorlinks=true,linkcolor=blue,citecolor=blue,urlcolor=blue]{hyperref}
\usepackage{txfonts}
\usepackage{dsfont}
\usepackage{enumitem}
\usepackage[utf8]{inputenc}
\usepackage{bbold}
\usepackage{mathtools}
\usepackage{lipsum, babel}
\usepackage[normalem]{ulem}
\usepackage{tikz}
\usepackage{pgfplots}
\usepackage{amsmath}
\pgfplotsset{compat=1.18}

\begin{document}

\title{
Observable Scaling Hierarchies in Multiphoton Dissipative Quantum Sensing} 
\author{Shahram Panahiyan$^{1,2}$}
\affiliation{$^1$ Max Planck Institute for the Structure and Dynamics of Matter, Luruper Chaussee 149, 22761 Hamburg, Germany }
\affiliation{$^2$ University of Hamburg, Luruper Chaussee 149, Hamburg D-22761, Germany}

\date{\today}

\begin{abstract}

We investigate how quantum correlations in squeezed driving fields determine scaling laws in dissipative multiphoton quantum sensing. Independently squeezed fields yield \emph{factorized} scaling, with separate absorption and emission contributions and nonlinear thresholds that suppress exponential scaling in linear processes. In contrast, jointly squeezed fields generate \emph{collective} scaling governed by the total nonlinear photon order of the dissipative interaction. Remarkably, we show that normally ordered observables do not inherit the full nonlinear scaling of the underlying multiphoton fluctuations. Instead, they exhibit asymptotic behavior with an effective nonlinear order reduced by one. This arises because normally ordered observables probe only part of the underlying multiphoton fluctuation structure. These findings reveal how multiphoton fluctuations, quantum correlations, and measurement structure jointly determine the experimentally accessible sensitivity of nonlinear dissipative quantum sensors and establish design principles for quantum sensing protocols based on structured squeezed light.

\end{abstract}

\maketitle

\section{Introduction} 

Dissipative quantum sensing has emerged as a powerful approach for probing weak physical parameters through environment-induced transitions and nonequilibrium light-matter interactions \cite{Montenegro2025,Raghunandan2018,Vaidya2025,Sarkar2025,Peng2024,Beaulieu2025}. Unlike purely coherent interferometric protocols, dissipative sensors can directly exploit loss, emission, and relaxation processes as metrological resources, enabling parameter estimation in open quantum systems where dissipation is unavoidable or even engineered \cite{Raghunandan2018,Xie2020,Harrington2022,DiCandia2023,Alushi2025,Beaulieu2025}. As dissipative control becomes increasingly accessible in photonic \cite{Parto2023}, atomic \cite{Diehl2008,Barreiro2011,Harrington2022,Ding2024}, and superconducting \cite{Fitzpatrick2017,Beaulieu2025} platforms, understanding the fundamental scaling laws governing dissipative quantum sensors has become an important problem in quantum metrology \cite{Alipour2014,Beaulieu2025}.

Multiphoton dissipative processes occupy a special role within this framework because their transition rates depend directly on high-order field correlations \cite{Schlawin2018,Panahiyan2023,Schlawin2026,kosheleva2025enhancedmultiphotonionizationdriven,Yu2021,Wolski2020,Yu2021,Reiter2017}. Unlike linear dissipative channels, multiphoton absorption and emission probe nonlinear fluctuation sectors of the driving fields, making them intrinsically sensitive to quantum correlations and nonequilibrium photon statistics. Engineered multiphoton dissipation therefore provides a natural setting for investigating how nonclassical fluctuations shape the scaling structure of open-system quantum sensing \cite{Giovannetti2011,Dorfman16,Polino2020,Casacio2021,Montenegro2025}.


Squeezed optical fields strongly amplify multiphoton fluctuations \cite{Lvovsky2015,Schnabel2017,Lawrie2019,Michael19,Atkinson2021,Kalash2026} and can substantially modify the response of nonlinear dissipative \cite{Schlawin2018,Panahiyan2023,Schlawin2026} and non-dissipative \cite{Spasibko2017,Michael19,Tzur2024,Rasputnyi2024,Rasputnyi2024N,Kondo2024,Akatev2026} channels. However, the metrological consequences of these enhanced fluctuations depend not only on their magnitude but also on their correlation structure and experimental accessibility. Independently squeezed optical modes generate fluctuations locally within each field, whereas jointly squeezed fields produce intermode correlations that directly couple absorption and emission processes. This raises a fundamental question: to what extent do the enhanced multiphoton fluctuations generated by squeezing translate into experimentally accessible metrological advantage? 
Since experimentally accessible observables probe only specific moments of the underlying multiphoton fluctuations, it remains unclear how distinct squeezing-induced correlation structures determine the observable scaling of nonlinear dissipative quantum sensors.


Previous studies have established important roles for squeezed multiphoton states in quantum metrology and for higher-order quantum correlations in nonlinear entanglement characterization. 
In particular, scalable multiphoton quantum metrology with two-mode squeezed vacuum states and photon-number-resolved detection has demonstrated the metrological value of exploiting complete multiphoton statistics in unitary sensing protocols, establishing that the choice of measurement determines the accessible metrological information \cite{You2021}. 
Related studies have developed hierarchies of higher-order entanglement witnesses and separability criteria for characterizing nonlinear quantum correlations \cite{Zhang2021,Li2022}. 
The present work addresses a complementary problem: rather than unitary parameter encoding or entanglement characterization, we consider multiphoton absorption-emission dynamics described by a Lindblad master equation and investigate how the correlation structure of independently and jointly squeezed fields determines the asymptotic scaling of experimentally accessible observables. 
Accordingly, the hierarchy developed here concerns dissipative metrological sensitivity rather than a hierarchy of entanglement witnesses or optimal measurements. The collective scaling considered here likewise refers to non-factorizable intermode contributions to experimentally accessible observables, rather than to many-body collective quantum states.


Within this Lindblad-based framework, we derive analytical scaling laws and show that dissipative multiphoton sensing exhibits two distinct asymptotic structures determined by the squeezing architecture of the driving fields. 
Independently squeezed pump and signal modes produce factorized scaling, in which absorption and emission contribute separately to the sensitivity, whereas jointly squeezed fields generate collective scaling governed by the total nonlinear photon order of the dissipative interaction. 
We further establish a universal one-order reduction: experimentally accessible normally ordered observables exhibit asymptotic scaling corresponding to a nonlinear order effectively reduced by one relative to the underlying multiphoton fluctuations. 
In jointly squeezed fields, intermode correlations modify this scaling structure through interference between local and collective fluctuation channels, leading to observable-dependent deviations and exceptional symmetry-induced regimes. 
Finally, we show that in this dissipative setting the structure of quantum correlations determines both the nonlinear response and the relative performance of measurements: independently squeezed fields exhibit a regime-dependent hierarchy with no globally optimal observable, while jointly squeezed fields lead to a constrained, asymmetry-dependent hierarchy of observable performance across different regimes.

These results reveal how squeezing correlation structure and measurement structure jointly constrain nonlinear quantum scaling in dissipative sensing, with direct implications for the design of optimal sensing protocols.

\begin{figure*}
\centering
\includegraphics[width=0.95\textwidth]{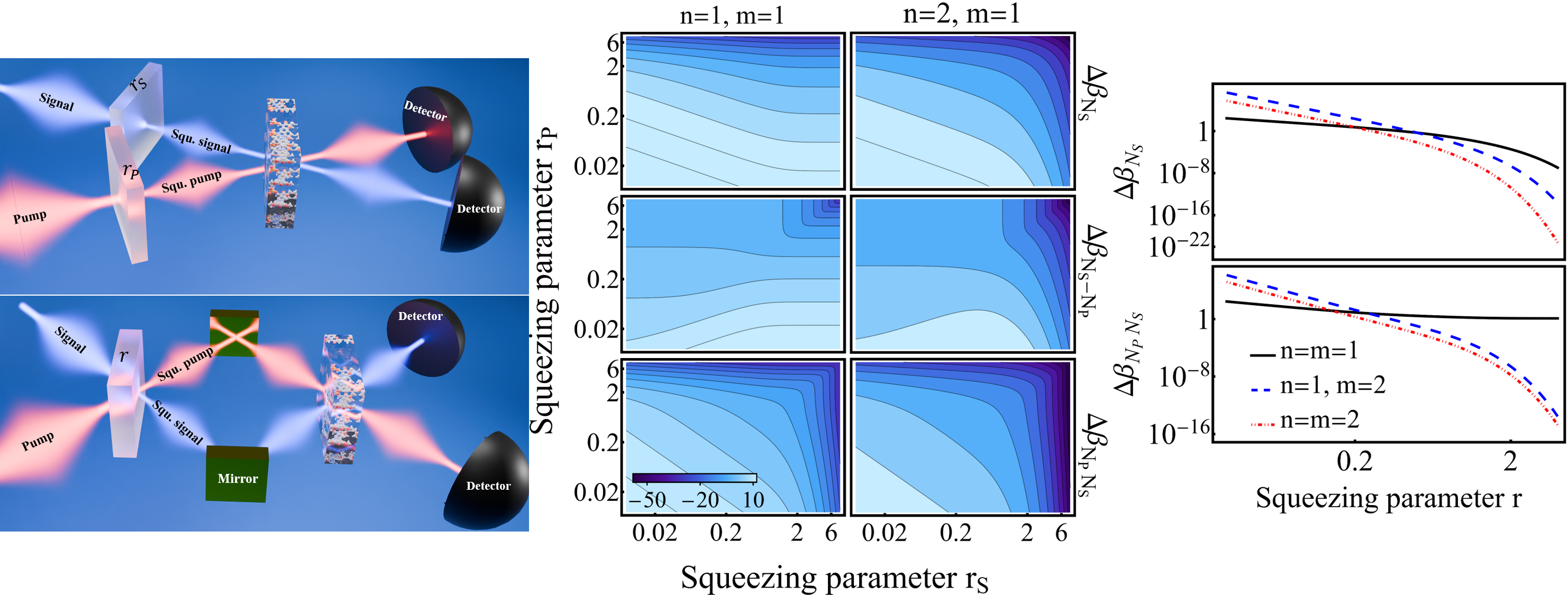}
\llap{\parbox[b]{13.68in}{(a)\\\rule{0ex}{2.3in}}} 
\llap{\parbox[b]{8.45in}{(b)\\\rule{0ex}{2.43in}}} 
\llap{\parbox[b]{3.4in}{(c)\\\rule{0ex}{2.2in}}} 
\caption{
(a) Schematic illustration of the two illumination configurations considered: independently squeezed pump and signal modes (top) and jointly squeezed fields (bottom).
(b) Sensitivity as a function of the squeezing parameters for independently squeezed fields for representative observables. Distinct regimes emerge, including saturation features associated with nonlinear thresholds, reflecting the factorized contributions of absorption and emission processes.
(c) Sensitivity as a function of a single squeezing parameter for jointly squeezed fields, for photon counting and photon-correlation measurements. Exponential scaling is generally observed in the high-gain regime, except for correlation measurements in the symmetric linear case, where the exponential dependence disappears, and the sensitivity saturates. 
}
\label{fig:relative}
\end{figure*}



\section{Multiphoton absorption and emission theory} 

We consider a dissipative multiphoton process in which $m$ pump photons are absorbed, and $n$ signal photons are emitted. The optical fields are modeled as two independent bosonic modes with annihilation operators $\hat{a}_P$ and $\hat{a}_S$, satisfying canonical commutation relations.
We assume weak coupling between the sample and the optical fields and a fast relaxation of the sample degrees of freedom. Under these conditions, the sample acts as a Markovian reservoir for the fields (see Appendix \ref{app:LindbladianJust}), allowing the internal dynamics of the sample to be adiabatically eliminated, resulting in the following Lindblad master equation
\begin{align}
\frac{d}{dt} \hat{\rho} = \beta \mathcal{L}[\hat{\rho}]
= \beta \Bigl( 2 \hat{a}_{P}^m (\hat{a}_{S}^{\dagger})^n \, \hat{\rho} \, (\hat{a}_{P}^{\dagger})^m \hat{a}_{S}^n 
- \bigl\{ \hat{\rho}, (\hat{a}_{P}^{\dagger})^m \hat{a}_{S}^n \hat{a}_{P}^m (\hat{a}_{S}^{\dagger})^n \bigr\} \Bigr),
\label{eq:Lindblad}
\end{align}
in which the jump operator is $\hat{a}_{P}^m (\hat{a}_{S}^{\dagger})^n$ and $\beta$ is the absorption-emission coefficient related to the total cross-section of the process. 

Such a Lindbladian description of multiphoton absorption and emission processes has a well-established microscopic basis in quantum optics. Higher-order perturbation theory relates multiphoton transition rates to high-order normally ordered field correlations \cite{Agarwal1970}, while density-matrix treatments provide master-equation descriptions of multiphoton absorption and emission processes \cite{Zubairy1980}. 
In particular, Sáinz de los Terreros et al. and Simaan developed master-equation descriptions of hyper-Raman processes, including the three-photon process in which two pump photons are absorbed and one signal photon is emitted \cite{Terreros1985,Simaan1978}, providing a concrete realization of the multiphoton absorption-emission structure considered here.
More generally, effective Lindblad operators can be obtained by adiabatically eliminating rapidly relaxing microscopic degrees of freedom \cite{Reiter2012}. Another microscopic realization particularly relevant to the present setting is stimulated Raman scattering, where the optical fields exchange energy with material vibrational modes. In a recent treatment of quantum-enhanced stimulated Raman scattering, Schlawin and Gessner showed that, when the pump-probe frequency difference is resonant with a material vibration, the reduced optical-field dynamics acquire a Lindbladian contribution describing this irreversible energy exchange \cite{Schlawin2026}.

To evaluate the expectation value of any operator in this framework, we consider \cite{Panahiyan2022,Panahiyan2023,shukla2025enhancingmeasurementprecisionnondegenerate}
\begin{align} 
\langle \hat{O} \rangle &= \text{tr} \left\{ \hat{O} e^{ \beta \mathcal{L} } e^{ \mathcal{L}_{squ.} } \rho_0 \right\}, \label{eq.General} 
\end{align}
where $\rho_0 = \vert 0,0 \rangle \langle 0,0 \vert$ is the two-mode vacuum, and 
$\mathcal{L}_{squ.}$ denotes the squeezing operation, performing
\begin{align}
e^{ \mathcal{L}_{squ.} }  \rho &\equiv \hat{U}_{k} \rho \hat{U}^{\dagger}_{k}. \label{eq.U_OPA}
\end{align}
The squeezing can be applied either (I) independently to each mode, with $\hat{U}_{k} = \exp \left[\zeta_k a^{\dagger 2}_{k}/2-h.c. \right]$ and $\zeta_k = r_k e^{i\phi_k}$ where $r_k$ and $\phi_k$ denote the gain parameter and phase of the mode and $k \in \{P,S\}$, or (II) jointly as a two-mode squeezing operation with $\hat{U}_{k} =\hat{U} =  \exp \left[ \zeta \hat{a}^{\dagger}_P \hat{a}^{\dagger}_S - h.c. \right]$, where $\zeta = r e^{i\phi}$. The squeezed fields are taken as externally prepared quantum resources; hence, the quantum correlations considered here are not generated by the nonlinear dissipative interaction or by any particular microscopic energy-level quantization. 

Considering that multiphoton absorption-emission processes have small cross sections \cite{Michael19,Min2025}, we focus on the weak-interaction limit, $\beta \ll 1$, where the Lindblad evolution can be linearized as $e^{ \beta \mathcal{L} } \simeq \mathbb{1}+\beta \mathcal{L}$, corresponding to a first-order perturbative expansion in $\beta$. In the Heisenberg picture, this yields input-output relations of the form $\hat{a}_k \to \hat{a}_k + \beta\, \mathcal{F}_k$ in which
\begin{align}
\mathcal{F}_S=\frac{d}{dt} \hat{a}_{S} &= \, n \, (\hat{a}_{P}^{\dagger})^m \hat{a}_{P}^m \hat{a}_{S}^n (\hat{a}_{S}^{\dagger})^{\, n-1},\\
\mathcal{F}_P=\frac{d}{dt} \hat{a}_{P} &= - \, m \, (\hat{a}_{P}^{\dagger})^{\, m-1} \hat{a}_{P}^m \hat{a}_{S}^n (\hat{a}_{S}^{\dagger})^{\, n},
\end{align}
where we have used the standard identities $[\hat{a}_{S}^n,\hat{a}_{S}^{\dagger}] = n \hat{a}_{S}^{\, n-1}$ 
and $[(\hat{a}_{S}^{\dagger})^n, \hat{a}_{S}] = -n (\hat{a}_{S}^{\dagger})^{\, n-1}$. Using these relations, we define the input-output transformations of the fields inside the sample
\begin{align}
\hat{a}_{S} &\rightarrow \hat{a}_{S}^{\prime} = \hat{a}_{S} + \beta \, n \, (\hat{a}_{P}^{\dagger})^m \hat{a}_{P}^m \hat{a}_{S}^n (\hat{a}_{S}^{\dagger})^{\, n-1}, \label{eq:transf_as} \\
\hat{a}_{P} &\rightarrow \hat{a}_{P}^{\prime} = \hat{a}_{P} - \beta \, m \, (\hat{a}_{P}^{\dagger})^{\, m-1} \hat{a}_{P}^m \hat{a}_{S}^n (\hat{a}_{S}^{\dagger})^n.   \label{eq:transf_apdg}
\end{align}
This formulation is particularly advantageous because it provides direct access to the time evolution of experimentally relevant observables.

We consider Gaussian input states obtained by applying squeezing procedures to the vacuum. The output fields are analyzed through experimentally accessible observables, including photon-number
\(\hat{N}_k = \hat{a}_k^\dagger \hat{a}_k\), their difference \(\hat{N}_S - \hat{N}_P\), and the photon-number correlation \(\hat{N}_P\hat{N}_S \), evaluated using normal-ordering techniques ~\cite{deepak2023,Meng2019, Meng2023} (see Appendix \ref{app:analyticalObs} for closed-form expressions).

The present effective description concerns the quantization of the optical fields and their nonlinear dissipative interaction, rather than microscopic mechanisms of energy-level quantization or spectral restructuring such as quantum confinement, avoided crossings, crystal-field splitting, or dressing-induced level shifts.


\section{Estimation strategy}

To quantify the precision in estimating \(\beta\), we employ the standard error-propagation formula \cite{Toth2014,Niezgoda2019}
\begin{align}
\Delta \beta^2 = \frac{\Delta \hat{O}^2}{\left| \partial \langle \hat{O} \rangle / \partial \beta \right|^2}, \label{eq:error}
\end{align}
where \(\Delta \hat{O}\) is the standard deviation. This approach directly relates the intrinsic quantum fluctuations of \(\hat{O}\) to its changes in \(\beta\), providing a transparent and experimentally relevant measure of estimation precision \cite{BachorRalph,HughesHase,Panahiyan2022,Panahiyan2023,Schlawin2026}. Since the observables considered here are directly measurable via standard photon-counting or coincidence-detection schemes, the error-propagation uncertainty offers a practical benchmark for achievable sensitivity in multiphoton absorption-emission experiments \cite{BachorRalph,HughesHase}. In what follows, the sensitivity is defined as the estimated uncertainty in \(\beta\), so smaller values of sensitivity correspond to higher precision.


\begin{table*}[t]
\centering
\begin{tabular}{|c|c|c|c|}
\hline
\textbf{Regimes of interest} 
& $\Delta \beta_{N_S}$ 
& $\Delta \beta_{N_S-N_P}$ 
& $\Delta \beta_{N_PN_S}$ \\
\hline
\multicolumn{4}{|c|}{\textit{Independent pump-signal squeezing}} \\
\hline
$r_P\leqslant 1$, $r_S\leqslant 1$ 
& $\mathcal{O}(r_S^2 r_P^{-4m})$ 
& $\mathcal{O}(r_P^{2-4m} + r_P^{-4m} r_S^2)$ 
& $\mathcal{O}(r_P^{2-4m} r_S^2 + r_P^{4-4m} r_S^4)$ \\
\hline
$r_P\leqslant 1$, $r_S\gg 1$ 
& $\mathcal{O}(e^{-4(n-1)r_S} r_P^{-4m})$ 
& $\mathcal{O}(r_P^{-4m} e^{-4(n-1) r_S})$ 
& $\mathcal{O}(r_P^{2-4m} e^{-4n r_S})$ \\
\hline
$r_P\gg 1$, $r_S\leqslant 1$ 
& $\mathcal{O}(r_S^2 e^{-4mr_P})$ 
& $\mathcal{O}(e^{-4 (m-1) r_P} r_S^2)$ 
& $\mathcal{O}(r_S^2 e^{-4m r_P})$ \\
\hline
$r_P\gg 1$, $r_S\gg 1$ 
& $\mathcal{O}(e^{-4(n-1)r_S} e^{-4mr_P})$ 
& $\mathcal{O}(e^{-4((m-1) r_P + n r_S)} + e^{-4(m r_P + (n-1) r_S)})$ 
& $\mathcal{O}\!\left(\frac{ e^{-4 ((m-1) r_P +(n-1) r_S )}} {(c(n,m) e^{2 r_P}  + c^{\prime}(n,m) e^{2 r_S})^2}\right)$ \\
\hline
\multicolumn{4}{|c|}{\textit{Joint pump-signal squeezing}} \\
\hline
$r \leqslant 1$ 
& $\mathcal{O}(r^{2-4 m})$ 
& --- 
& $\mathcal{O}(r^{2-4 m})$ \\
\hline
$r \gg 1$ 
& $\mathcal{O}(e^{-4(m+n-1) r})$ 
& --- 
& $\mathcal{O}(e^{-4(m+n-1) r})$ and $\mathcal{O}(e^{-8(n-1) r}) \text{ if } n=m$ \\
\hline
\end{tabular}
\caption{Leading-order asymptotic scaling of the sensitivity measures $\Delta \beta_{N_S}$, $\Delta \beta_{N_S-N_P}$, and $\Delta \beta_{N_P N_S}$ for independently and jointly squeezed pump and signal fields. 
For both cases, power-law (algebraic) behavior appears when a field remains in the low-gain regime, while exponential suppression emerges once the corresponding field enters the high-gain regime. The decay rates are governed by the nonlinear emission and absorption orders $n$ and $m$, respectively. In the independently squeezed case, cross-exponentials in the high-gain regimes capture the interplay of multiphoton emission and absorption. 
In the joint-squeezing scenario, the scaling reduces to a cooperative exponential decay in the high-gain regime. Here, $c(n,m)$ and $c^{\prime}(n,m)$ are constant prefactors given in Eq. \eqref{eq:sepcor}.}
\label{tab:fullscal_combined}
\end{table*}

\section{Observable scaling hierarchy under multiphoton dissipation}


We now establish the structural principles governing the asymptotic scaling of multiphoton dissipative sensing. 

\textbf{Theorem (Universal one-order reduction).}
The asymptotic scaling of dissipative sensitivities in an $m$-photon absorption and $n$-photon emission process is determined by the highest-order normally ordered field moments contributing to the measured observable.

For dissipative multiphoton processes generated by
$\hat L=\hat a_P^m(\hat a_S^\dagger)^n$, the underlying multiphoton fluctuations scale with the full nonlinear order ($m+n$) of the dissipative interaction. For normally ordered observables, however, the leading observable scaling is reduced by one nonlinear order relative to the underlying multiphoton fluctuations, provided the dominant fluctuation contributions remain nonvanishing. Physically, this reduction reflects the fact that normally ordered observables probe only part of the underlying multiphoton fluctuation structure. 


\textbf{Proposition (Correlation-induced scaling modification).}
For jointly squeezed fields, intermode correlations generate interference between local and collective fluctuation contributions to the observable moments. These interference effects can modify or suppress the leading asymptotic scaling for specific observables and symmetry configurations, producing observable-dependent deviations from the universal one-order-reduced scaling structure.


The explicit observable realizations of these scalings are derived through the asymptotic analyses presented in the following section, and the proof follows from the exhaustive analysis of Appendix \ref{sec:scaling_study} for the stated observable class.

\begin{figure*}
\centering
\includegraphics[width=0.8\textwidth]{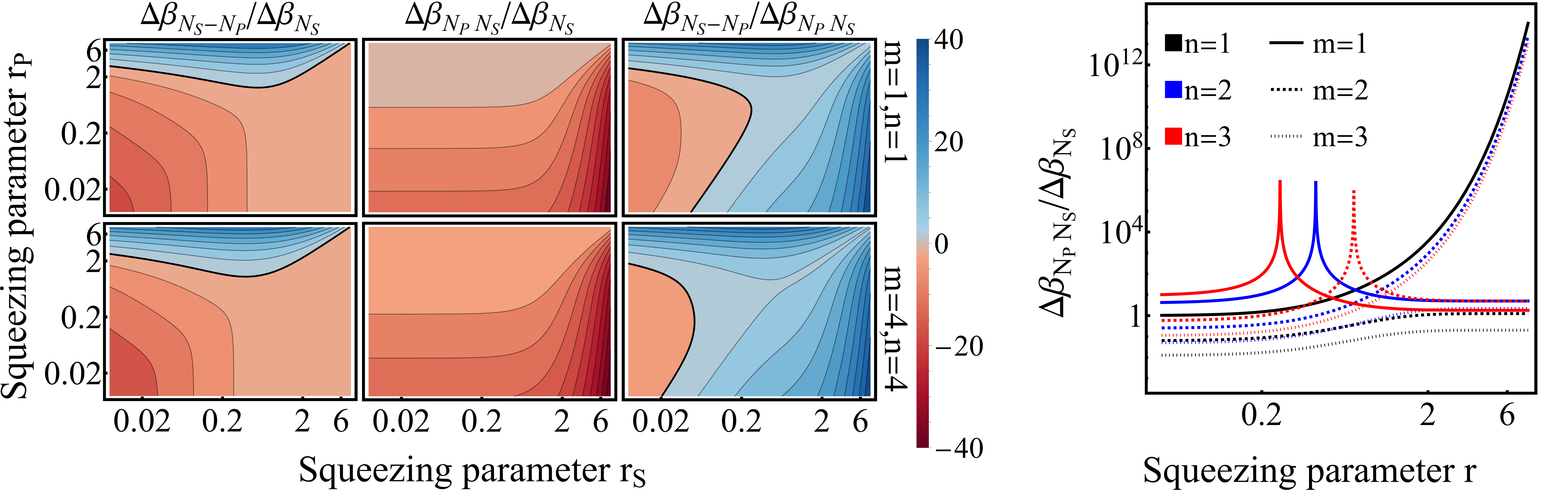}
\llap{\parbox[b]{11in}{(a)\\\rule{0ex}{1.7in}}} 
\llap{\parbox[b]{3.5in}{(b)\\\rule{0ex}{1.7in}}} 
\caption{
Relative performance of measurement strategies for independently and jointly squeezed illumination. 
(a) Logarithmic ratio $\Delta \beta_{N_S-N_P}/\Delta \beta_{N_S}$ (left panel), $\Delta \beta_{N_P N_S}/\Delta \beta_{N_S}$ (middle panel), and $\Delta \beta_{N_S-N_P}/\Delta \beta_{N_P N_S}$ (right panel) as a function of $r_S$ and $r_P$ for different nonlinear orders $(m,n)$.  The ratio $\Delta \beta_{N_P N_S}/\Delta \beta_{N_S}$ remains below unity across the entire parameter space, while $\Delta \beta_{N_S-N_P}/\Delta \beta_{N_S}$ and $\Delta \beta_{N_S-N_P}/\Delta \beta_{N_P N_S}$ exhibit a regime-dependent behavior revealing nontrivial competitions between different observables and correlation observable. Across all panels, the qualitative structure is preserved for different $(m,n)$, with the nonlinear orders primarily shifting the boundaries (denoted by dark solid line) between regimes. 
(b) Ratio $\Delta \beta_{N_P N_S}/\Delta \beta_{N_S}$ for jointly squeezed fields as a function of the squeezing parameter $r$ in logarithmic scaling. The ratio shows mixed behavior determined by the gain parameter and order parameter. 
}
\label{fig:relative}
\end{figure*}

\section{Scaling behavior and physical interpretation} \label{sec:scaling}

We first consider independently squeezed pump and signal modes in which the nonclassical fluctuations of each field can be tuned separately, allowing us to isolate how single-mode squeezing modifies multiphoton absorption and emission processes. The input-output transformations before the interaction with the sample are
$\hat{a}_{k} \rightarrow \cosh r_k\, \hat{a}_{k} + e^{i \phi_{k}} \sinh r_k\, \hat{a}_{k}^{\dagger}$
\cite{BarnettRadmore,Panahiyan2022}. We next consider jointly squeezed pump and signal fields prepared in a two-mode squeezed state in which quantum correlations directly couple the two modes, thereby modifying the joint photon statistics that enter multiphoton transition amplitudes. The corresponding transformations are
$\hat{a}_{k} \rightarrow \cosh r\, \hat{a}_{k} + e^{i \phi} \sinh r\, \hat{a}_{j}^{\dagger}$
with $j\in \{P,S\}$ and $j \ne k$ \cite{BarnettRadmore,shukla2025enhancingmeasurementprecisionnondegenerate}. Closed-form expressions for the observables are given in 
Appendices \ref{sec:observablesSep} and \ref{sec:observablesJoin}.

We analyze the scaling of sensitivities for independently and jointly squeezed fields across low- and high-gain regimes, with the results given in Appendix \ref{sec:scaling_study}, and their scaling behaviors summarized in Table \ref{tab:fullscal_combined}. 

Two qualitatively distinct behaviors emerge. In the low-gain regime, the dynamics are pump-dominated since multiphoton absorption requires photons in the pump mode, while emission processes can proceed from vacuum fluctuations. As a result, the leading scaling is governed by the statistics of the pump field, with the signal contributing only subleading corrections. This asymmetry gives rise to algebraic scaling laws characteristic of a perturbative regime.
In the high-gain regime, exponential factors appear whose decay rates are determined by the nonlinear absorption and emission orders. In mixed regimes, algebraic and exponential contributions coexist, reflecting the interplay between weakly and strongly squeezed modes.

For independently squeezed fields, the high-gain scaling factorizes into contributions associated with the two modes, reflecting the action of the Lindblad generator through separate dissipative channels associated with the pump and signal modes. Exponential contributions from absorption and emission processes, therefore, appear independently, and the detailed scaling remains observable-dependent. A direct consequence is the emergence of \emph{nonlinear thresholds}, whereby the exponential scaling disappears when the corresponding absorption or emission process becomes linear (e.g., linear emission in photon number measurement). Independent squeezing thus probes absorption and emission pathways separately, revealing their individual nonlinear character.

In contrast, jointly squeezed illumination leads to the scaling reorganizing into a collective form governed by a single squeezing parameter $r$. This is because intermode correlations reorganize the photon statistics and activate the collective structure of the multiphoton jump operator. The sensing dynamics then behave as a single cooperative multiphoton channel.
Consequently, in the high-gain regime, photon-counting and correlation-based observables generally exhibit scaling in which exponential suppression is controlled by the total nonlinear order of the multiphoton interaction, $(m+n-1)$. 
This collective behavior, however, is observable-dependent and therefore not universally accessible at the level of measurable sensitivities. Its manifestation depends on how the chosen observable samples the underlying quantum fluctuations. In particular, symmetry constraints and fluctuation properties of the measurement operator can suppress the expected exponential scaling, leading in some cases to saturation in the high-gain regime (e.g., linear processes with $m=n=1$ for photon correlations) or to ill-defined sensitivities (e.g., the photon-number difference, whose variance vanishes under two-mode squeezing).

In the high-gain equal-squeezing limit ($r_P = r_S \gg 1$), the leading exponential scaling for independently and jointly squeezed fields coincides, yielding an effective universal behavior governed by the total nonlinear order $(m + n - 1)$. This realizes the universal one-order-reduction theorem established above.
In this regime, the system behaves as a single nonlinear dissipative channel, where the combined multiphoton absorption-emission process determines the asymptotic metrological enhancement. Subleading terms and observable-dependent prefactors, however, retain signatures of the underlying squeezing configuration.
We quantify this by analyzing the relative performance of experimentally accessible measurements in the next section.

We emphasize that this reduction is a property of the asymptotic error-propagation sensitivities under nonlinear Lindblad dynamics, rather than a statement about optimal information extraction: it does not follow from the Fisher-information analyses of unitary multiphoton metrology \cite{You2021}, where different detection operators are known to extract different amounts of information from a given state, but instead originates from the operator structure of the dissipative multiphoton process itself. In particular, the $\beta$-derivative of a normally ordered observable involves field correlations of one order lower than the fluctuations generated by the full $(m + n)$-photon interaction, which is the origin of the one-order-reduced asymptotic scaling.

This behavior admits a natural interpretation as a nonlinear dissipative interferometer, in which multiphoton processes amplify the effective squeezing resource by reducing the number of independent fluctuation channels. The achievable precision, therefore, scales exponentially with the effective nonlinear order, establishing a general design principle for dissipative quantum sensors: the attainable scaling is set by the total photon number participating in the interaction, while its accessibility depends on both the entanglement structure of the fields and the measurement used to extract the signal.

\section{Relative performance of measurement strategies}

To assess the relative performance of different measurement strategies, we present Figure~\ref{fig:relative}, which summarizes these comparisons in terms of logarithmic ratios of sensitivities across the relevant parameter regimes.


For independent squeezing, correlation-based measurements provide a robust metrological advantage over photon counting across the entire $(r_P,r_S)$ parameter space for all nonlinear orders considered in this configuration (see Figs.~\ref{fig:relative}(a) and \ref{fig:relativeA} in the Appendix). 
In contrast, the ratio $\Delta \beta_{N_S-N_P}/\Delta \beta_{N_S}$ displays both enhancement and degradation regions across the $(r_P,r_S)$ plane. In particular, it exceeds unity for a high-gain pump field, whereas it remains below unity in the complementary regime. A similar competition is observed in Fig.~\ref{fig:relative}(a) for the ratio $\Delta \beta_{N_S-N_P}/\Delta \beta_{N_P N_S}$, which is smaller than unity for weak squeezing and becomes larger than unity as both $r_P$ and $r_S$ increase. 
Notably, the qualitative behaviors detailed above are largely preserved across different nonlinear orders, with $(m,n)$ primarily affecting the extent and location of the boundaries between advantage and disadvantage regions (see Fig. \ref{fig:relativeA} in the Appendix).

For jointly squeezed fields, the relative performance of photon-counting and correlation measurements exhibits a nontrivial asymptotic dependence on the nonlinear asymmetry of the dissipative interaction. As shown in Fig.~\ref{fig:relative}(b) (and Fig. \ref{fig:relativeAA} in the Appendix), correlation measurements outperform photon counting in the high-gain regime whenever the interaction is sufficiently asymmetric, $m-n\ge 2$ or $n-m \ge 3$, for which the sensitivity ratio falls below unity asymptotically. In the remaining cases, including symmetric and weakly asymmetric processes, the hierarchy reverses at large gain, and photon counting provides superior sensitivity. Moreover, for $m-n\ge 2$, the correlation observable remains advantageous across all gain regimes. These results show that the observable hierarchy in jointly squeezed dissipative sensing depends nontrivially on both the magnitude and direction of the nonlinear asymmetry.
At high gain, the ratios $\Delta \beta_{N_P N_S}/\Delta \beta_{N_S}$ for $n\neq m$ saturate, signaling the onset of the asymptotic scaling regime discussed previously. 


Taken together, these results show that, while independent squeezing admits a broadly flexible observable hierarchy, jointly squeezed fields produce a more constrained asymmetry-dependent structure governed by collective multiphoton correlations and measurement accessibility. 

We note that the hierarchy identified here is complementary to measurement-optimality analyses in unitary multiphoton metrology, where photon-number-resolved detection has been shown to maximize the accessible information by exploiting the complete multiphoton statistics \cite{You2021}. Rather than identifying a globally optimal measurement, the present analysis compares the asymptotic sensitivities of experimentally standard observables under dissipative dynamics, showing that their relative performance is governed by the nonlinear interaction orders and the correlation structure of the illumination, with no single observable being optimal across all regimes.

\section{Experimental realization}

The scaling hierarchy studied here can be investigated using established squeezed-light and photon-counting techniques. Independently squeezed pump and signal fields can be generated using separate single-mode squeezing processes \cite{Park2024}, while jointly squeezed fields can be produced through two-mode parametric interactions \cite{Guerrero2020,Kalash2026}. The photon-number observables considered here can be accessed using photon-number-resolving and coincidence-detection measurements \cite{Yu2021,Los2024}. The remaining requirement is a nonlinear dissipative optical channel realizing an effective multiphoton absorption-emission process; stimulated Raman scattering and hyper-Raman processes provide experimentally demonstrated examples of nonlinear light-matter interactions of this type \cite{Jones2024,Akatev2026}, while engineered dissipative light-matter interactions offer additional routes to effective nonlinear channels. Thus, the predicted factorized and collective scaling regimes and the associated observable hierarchy can, in principle, be tested by varying the squeezing gain and measuring the resulting photon-number statistics.

\section{Conclusion}

We have established the scaling structure governing multiphoton dissipative quantum sensing with squeezed light. While squeezed fields exponentially enhance the underlying multiphoton fluctuations, we showed that experimentally accessible sensitivities obey a universal one-order-reduced hierarchy determined by the structure of measurable observables. Independently and jointly squeezed fields realize distinct factorized and collective scaling regimes, with intermode correlations generating observable-dependent interference effects and exceptional symmetry-induced behavior.

Our results identify observable accessibility as a fundamental constraint in nonlinear dissipative quantum metrology, establish a unified framework for understanding how quantum correlations shape experimentally accessible scaling laws in open quantum systems, and provide general design principles for dissipative quantum sensors.

\section{Acknowledgments}
SP acknowledges support from the Hamburg Quantum Computing Initiative (HQIC) project EFRE. 
The EFRE project is co-financed by ERDF of the European Union and by “Fonds of the Hamburg Ministry of Science, Research, Equalities and Districts (BWFGB)”.

\bibliography{bibliography_photons}

\clearpage
\newpage

\appendix
\begin{widetext}

\renewcommand{\thesection}{\Alph{section}}
\numberwithin{equation}{section}
\renewcommand{\thefigure}{A.\arabic{figure}}
\setcounter{figure}{0}

\section{Justification for Lindbladian approach} \label{app:LindbladianJust}

The effective Lindblad description in Eq. (1) can be understood as the reduced dynamics of the optical fields after eliminating microscopic degrees of freedom of the sample. In a microscopic light-matter description, higher-order interactions generate transition amplitudes involving products of optical creation and annihilation operators. An $m$-photon absorption process gives rise to factors of $\hat a_P^m$, while the simultaneous emission of $n$ signal photons gives factors of $(\hat a_S^\dagger)^n$. Such multiphoton processes have been described microscopically using higher-order perturbation theory and density-matrix methods \cite{Agarwal1970,Zubairy1980}. 
In particular, the three-photon hyper-Raman process considered by Sáinz de los Terreros et al. \cite{Terreros1985} provides an explicit example involving the absorption of two pump photons and the emission of one signal photon. More generally, when the relevant microscopic degrees of freedom of the sample are rapidly relaxing, they can be adiabatically eliminated, resulting in effective dissipative operators acting on the optical fields \cite{Reiter2012}.

The corresponding multiphoton channel is constrained by energy conservation in the underlying microscopic process. If the sample undergoes a transition with energy change ($\Delta E_{\mathrm{sample}}$), the allowed $m$-pump/$n$-signal process satisfies the resonance condition $m\hbar\omega_P-n\hbar\omega_S\simeq \Delta E_{\mathrm{sample}}$,
where the equality is understood within the linewidth of the relevant transition. For example, in stimulated Raman scattering, a pump photon is converted into a signal photon while the sample undergoes a vibrational transition, giving $\omega_P-\omega_S\simeq\omega_{\mathrm{vib}}$.
This is the microscopic energy-exchange mechanism underlying Raman processes and is also the situation considered in the recent microscopic treatment of quantum-enhanced stimulated Raman scattering by Schlawin and Gessner \cite{Schlawin2026}. Thus, $m$ and $n$ are not arbitrary photon numbers imposed independently of the physical system; rather, they specify a particular nonlinear optical channel selected by the microscopic transition and its resonance condition. After the sample degrees of freedom are traced out, their energy exchange with the optical fields is encoded in the effective dissipative dynamics of Eq. (1).

The weak-coupling and short-memory assumptions underpinning the Markovian approximation are well justified in our study because the sample relaxation time is much shorter than the timescale over which the light fields evolve. Similarly, any coherent back-action of the sample on the fields is negligible due to the low probability of multiphoton absorption and emission events under the weak-interaction conditions assumed here. Saturation and depletion of the sample states can likewise be neglected in the weak-interaction regime assumed here, where the sample remains far from population inversion or significant occupation of excited states. Under these conditions, the sample acts as an effectively Markovian reservoir for the optical fields, and the resulting reduced dynamics are accurately described by the Lindblad master equation in Eq. (1).

\section{Closed-form of the Observables} \label{app:analyticalObs}

In this section, we present the explicit analytical expressions for the observables employed throughout this work. To evaluate the expectation values of the observables introduced in the main text, we first establish a set of operator identities that allow us to systematically normal order products of creation and annihilation operators. This step is essential because squeezed states and multiphoton processes naturally generate operator expressions involving high powers of noncommuting field operators. We employ modified binomial expansions for bosonic operators~\cite{deepak2023},
\begin{align}
(\hat{a}x+ \hat{a}^{\dagger}y)^n &=
\sum_{m=0}^{\lfloor n/2 \rfloor} \frac{n!(xy)^m}{m!(n-2m)!2^m}
\sum_{r=0}^{n-2m} \binom{n-2m}{r} x^r y^{n-2m-r}
(\hat{a}^{\dagger})^{n-2m-r} \hat{a}^r ,
\end{align}
which explicitly account for the noncommutativity of $\hat{a}$ and $\hat{a}^{\dagger}$. Acting on the vacuum state, this expansion simplifies to
\begin{align}
(\hat{a}x+ \hat{a}^{\dagger}y)^n |0\rangle
= n! \sum_{m=0}^{\lfloor n/2 \rfloor}
\frac{(\hat{a}^{\dagger})^{n-2m}}{m!(n-2m)!2^m}
x^m y^{n-m} |0\rangle ,
\end{align}
which is particularly useful for computing expectation values in squeezed-vacuum and related nonclassical states. To reorder products of creation and annihilation operators, we further use the identities
\begin{align}
\hat{a}^j (\hat{a}^{\dagger})^{k} &=
\sum_{s=0}^{\min(j,k)} \frac{j! k!}{s!(j-s)!(k-s)!}
(\hat{a}^{\dagger})^{k-s} \hat{a}^{j-s},
\\
(\hat{a}^{\dagger})^{j} \hat{a}^k &=
\sum_{s=0}^{\min(j,k)} \frac{j! k!}{s!(j-s)!(k-s)!}
(-1)^s \hat{a}^{k-s} (\hat{a}^{\dagger})^{j-s},
\end{align}
with additional related identities given in Refs.~\cite{Meng2019,Meng2023}. Using the above relations, we obtain the general vacuum expectation value
\begin{align}
\langle 0| (\hat{a}x+ \hat{a}^{\dagger}y)^n (\hat{a}w+ \hat{a}^{\dagger}z)^n |0\rangle
= \sum_{m=0}^{\lfloor n/2 \rfloor}
\frac{(n!)^2}{(m!)^2 (n-2m)! 2^{2m}}
\, w^{m} z^{\,n-m} x^{\,n-m} y^{m},
\label{eq.GeneralExp}
\end{align}
which serves as a central building block for evaluating photon number moments, differences, 
and correlations of the squeezed fields considered in this work. 

Working in the Heisenberg picture, we apply the input-output transformations derived from the Lindbladian formulation, given in Eqs.~\eqref{eq:transf_as} and \eqref{eq:transf_apdg}. These relations describe the effective evolution of the field operators induced by the multiphoton absorption-emission process. Using these transformations, we obtain closed-form expressions for the relevant photon number moments, differences, and correlations, which are listed below 

\begin{align}
\hat{N}_{S}&=
\hat{n}_{S}+2 n \beta (\hat{a}_{P}^{\dagger})^m \hat{a}_{P}^m \bigg[ \hat{a}_{S}^{n}  (\hat{a}_{S}^{\dagger})^{n} -n \hat{a}_{S}^{n-1}  (\hat{a}_{S}^{\dagger})^{n-1} \bigg]+\mathcal{O}(\beta^2), \label{eq.GeneralN}
\end{align}

\begin{align}
\hat{N}_{P}&=(\hat{a}_{P}^{\prime})^{\dagger}\hat{a}_{P}^{\prime}=\hat{n}_{P}-2 \beta m (\hat{a}_{P}^{\dagger})^m \hat{a}_{P}^m  \hat{a}_{S}^{n}  (\hat{a}_{S}^{\dagger})^{n}+\mathcal{O}(\beta^2), 
\label{eq.GeneralN}
\end{align}

\begin{align}
\hat{N}_{S}^2&=((\hat{a}_{S}^{\prime})^{\dagger}\hat{a}_{S}^{\prime})^2=\hat{n}_{S}^2+4n\beta (\hat{a}_{P}^{\dagger})^m \hat{a}_{P}^m \bigg[ \hat{a}_{S}^{n+1}  (\hat{a}_{S}^{\dagger})^{n+1} -(2n+1) \hat{a}_{S}^{n}  (\hat{a}_{S}^{\dagger})^{n} +n^2 \hat{a}_{S}^{n-1}  (\hat{a}_{S}^{\dagger})^{n-1} \bigg] +\mathcal{O}(\beta^2), \label{eq.GeneralN2}
\end{align}

\begin{align}
\hat{N}_{P}^2&=((\hat{a}_{P}^{\prime})^{\dagger}\hat{a}_{P}^{\prime})^2=\hat{n}_{P}^2-4\beta m \bigg[ (\hat{a}_{P}^{\dagger})^{m+1} \hat{a}_{P}^{m+1}+m (\hat{a}_{P}^{\dagger})^{m} \hat{a}_{P}^{m} \bigg] \hat{a}_{S}^{n}  (\hat{a}_{S}^{\dagger})^{n}+O(\beta^2), \label{eq.GeneralN2}
\end{align}

\begin{align}
\hat{N}_{P}\hat{N}_{S}&=(\hat{a}_{P}^{\prime})^{\dagger}\hat{a}_{P}^{\prime}(\hat{a}_{S}^{\prime})^{\dagger}\hat{a}_{S}^{\prime}=\hat{n}_{P}\hat{n}_{S}+2 \beta \bigg[ \bigg( n (\hat{a}_{P}^{\dagger})^{m+1} \hat{a}_{P}^{m+1} + n m (\hat{a}_{P}^{\dagger})^{m} \hat{a}_{P}^{m} \bigg) \bigg( \hat{a}_{S}^{n}(\hat{a}_{S}^{\dagger})^{n}-n\hat{a}_{S}^{n-1} (\hat{a}_{S}^{\dagger})^{n-1} \bigg)-
\notag
\\&
m (\hat{a}_{P}^{\dagger})^{m} \hat{a}_{P}^{m} \bigg(\hat{a}_{S}^{n+1} (\hat{a}_{S}^{\dagger})^{n+1}-(n+1)\hat{a}_{S}^{n} (\hat{a}_{S}^{\dagger})^{n} \bigg) \bigg]+ O(\beta^2),
\end{align}


\begin{align}
(\hat{N}_{P}\hat{N}_{S})^2&=((\hat{a}_{P}^{\prime})^{\dagger}\hat{a}_{P}^{\prime}(\hat{a}_{S}^{\prime})^{\dagger}\hat{a}_{S}^{\prime})^2=(\hat{n}_{P}\hat{n}_{S})^2+ \notag
\\&
4\beta n\bigg( (\hat{a}_{P}^{\dagger})^{m+2} \hat{a}_{P}^{m+2}+(2m+1) (\hat{a}_{P}^{\dagger})^{m+1}\hat{a}_{P}^{m+1}+m(\hat{a}_{P}^{\dagger})^{m}\hat{a}_{P}^{m} \bigg) \bigg(\hat{a}_{S}^{n+1}(\hat{a}_{S}^{\dagger})^{n+1}-(2n+1)\hat{a}_{S}^{n}(\hat{a}_{S}^{\dagger})^{n}+n^2 \hat{a}_{S}^{n-1} (\hat{a}_{S}^{\dagger})^{n-1}\bigg)- \notag
\\&
4\beta m\bigg( (\hat{a}_{P}^{\dagger})^{m+1}\hat{a}_{P}^{m+1}+m(\hat{a}_{P}^{\dagger})^{m}\hat{a}_{P}^{m} \bigg)\bigg( \hat{a}_{S}^{n+2}(\hat{a}_{S}^{\dagger})^{n+2}-(2n+3) \hat{a}_{S}^{n+1}(\hat{a}_{S}^{\dagger})^{n+1} +(n+1)^2\hat{a}_{S}^{n}(\hat{a}_{S}^{\dagger})^{n} \bigg)+\mathcal{O}(\beta^2),
\end{align}

\begin{align}
(\hat{N}_{S}-\hat{N}_{P})^2&=((\hat{a}_{S}^{\prime})^{\dagger}\hat{a}_{S}^{\prime}-(\hat{a}_{P}^{\prime})^{\dagger}\hat{a}_{P}^{\prime})^2= ((\hat{a}_{S}^{\prime})^{\dagger}\hat{a}_{S}^{\prime})^2+((\hat{a}_{P}^{\prime})^{\dagger}\hat{a}_{P}^{\prime})^2-(\hat{a}_{S}^{\prime})^{\dagger}\hat{a}_{S}^{\prime}(\hat{a}_{P}^{\prime})^{\dagger}\hat{a}_{P}^{\prime}-(\hat{a}_{P}^{\prime})^{\dagger}\hat{a}_{P}^{\prime}(\hat{a}_{S}^{\prime})^{\dagger}\hat{a}_{S}^{\prime},
\end{align}
in which 

\begin{align}
\hat{N}_{S}\hat{N}_{P}&=(\hat{a}_{S}^{\prime})^{\dagger}\hat{a}_{S}^{\prime}(\hat{a}_{P}^{\prime})^{\dagger}\hat{a}_{P}^{\prime}= \hat{n}_{S}\hat{n}_{P}+2 \beta\bigg[
\bigg(n(\hat{a}_{P}^{\dagger})^{m+1} \hat{a}_{P}^{m+1} +n m (\hat{a}_{P}^{\dagger})^{m} \hat{a}_{P}^{m} \bigg) \bigg(\hat{a}_{S}^{n} (\hat{a}_{S}^{\dagger})^{n}-n \hat{a}_{S}^{n-1} (\hat{a}_{S}^{\dagger})^{n-1} \bigg)-
 \notag
\\&
m \beta (\hat{a}_{P}^{\dagger})^{m} \hat{a}_{P}^{m} \bigg(\hat{a}_{S}^{n+1} (\hat{a}_{S}^{\dagger})^{n+1}-(n+1)\hat{a}_{S}^{n} (\hat{a}_{S}^{\dagger})^{n} \bigg)
\bigg]+ \mathcal{O}(\beta^2).
\label{eq.GeneralNSNP}
\end{align}

\begin{figure*}
\centering
\includegraphics[width=0.5\textwidth]{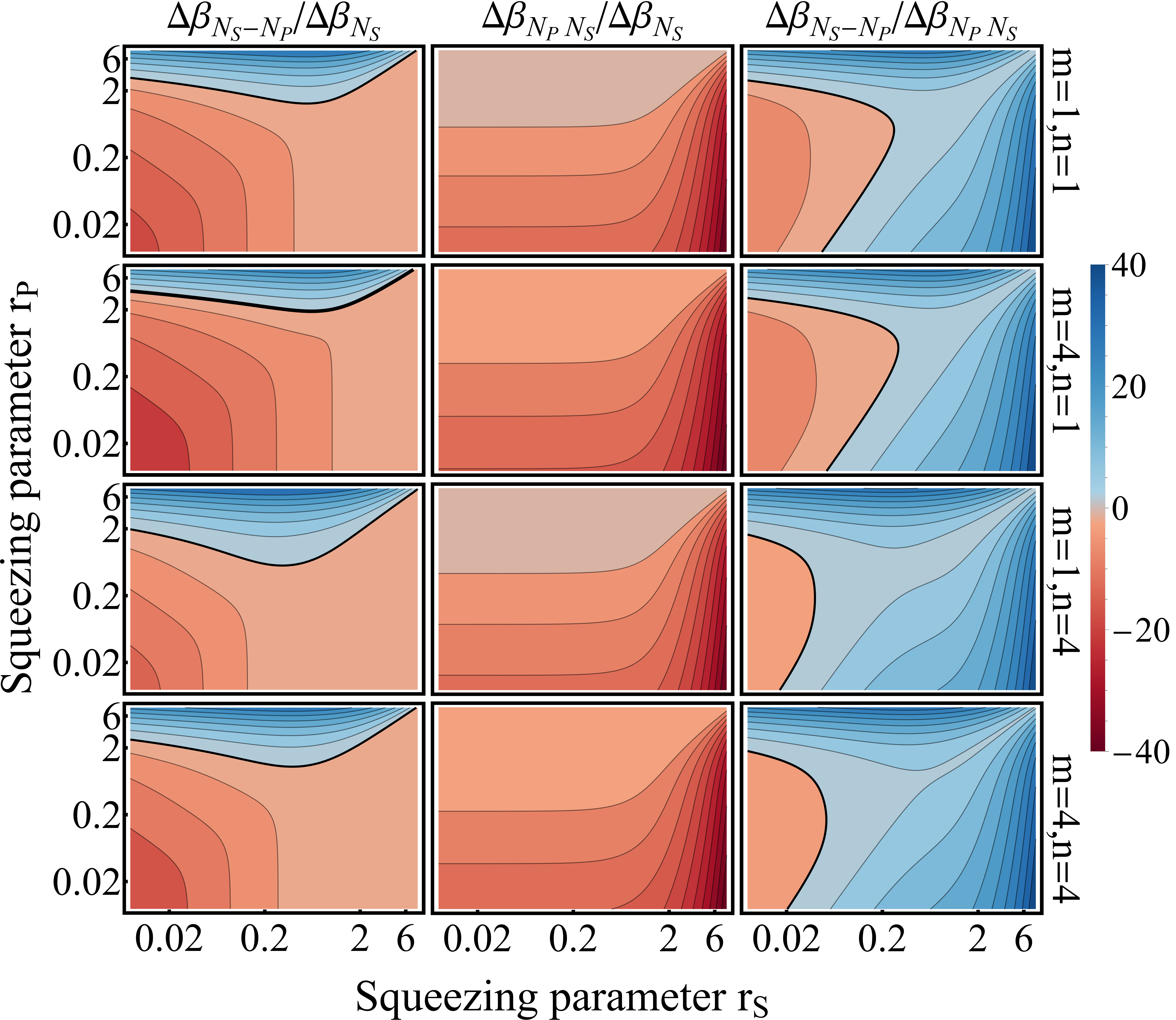}
\caption{
Relative performance of measurement strategies for independently squeezed illumination. 
}
\label{fig:relativeA}
\end{figure*}

\begin{figure*}
\centering
\includegraphics[width=0.5\textwidth]{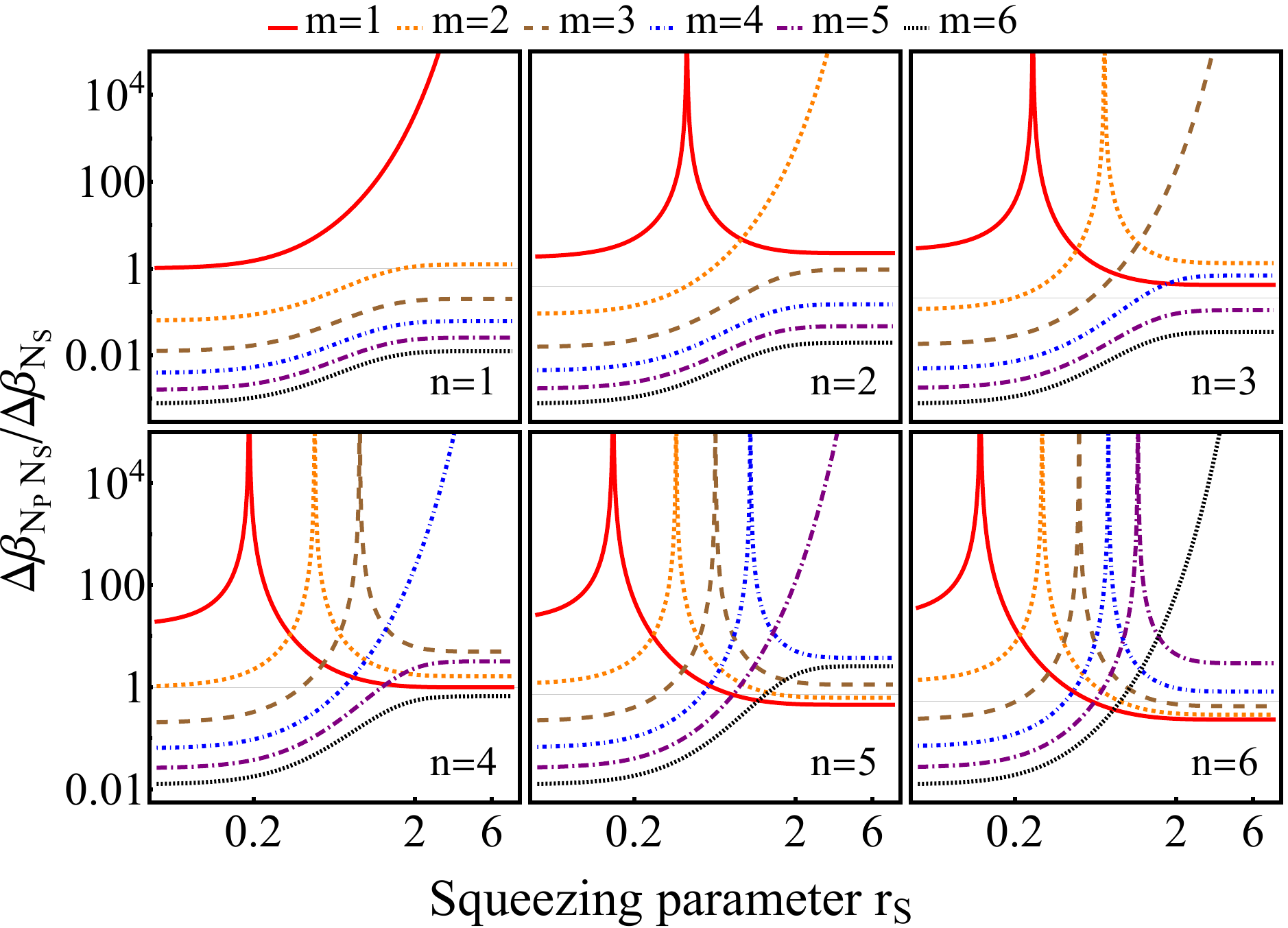}
\caption{
Relative performance of measurement strategies for jointly squeezed illumination. 
}
\label{fig:relativeAA}
\end{figure*}

\section{Calculated quantities for the case of independent squeezing} \label{sec:observablesSep}

In this section, we present the results for photon counting and its higher-order moments 
for the case in which the pump and signal fields are squeezed independently. In this configuration, the nonclassical properties of each field can be controlled separately, allowing us to isolate the influence of single-mode squeezing on multiphoton absorption and emission processes.

To evaluate the expectation values of the normally and antinormally ordered expressions appearing in Eqs.~\eqref{eq.GeneralN}-\eqref{eq.GeneralNSNP}, we make use of the general identity given in Eq.~\eqref{eq.GeneralExp}. These expectation values naturally arise when expressing multiphoton observables in terms of squeezed-field operators and encode the dependence of the nonlinear response on squeezing strength and phase. For the choice $x = e^{-i\phi}\sinh r$, $y = \cosh r$, $w = \cosh r$, and $z = e^{i\phi}\sinh r$, corresponding to normally ordered operator products, we obtain
\begin{align}
\langle 0| (\hat{a}x + \hat{a}^{\dagger}y)^n (\hat{a}w + \hat{a}^{\dagger}z)^n |0\rangle
= \sum_{m=0}^{\lfloor n/2 \rfloor}
\frac{(n!)^2}{(m!)^2 (n-2m)! \, 2^{2m}}
(\cosh r)^{2m} (\sinh r)^{2n-2m}.
\end{align}
Conversely, for $x = \cosh r$, $y = e^{i\phi}\sinh r$, $w = e^{-i\phi}\sinh r$, and $z = \cosh r$, which corresponds to antinormally ordered operator products, we find
\begin{align}
\langle 0| (\hat{a}x + \hat{a}^{\dagger}y)^n (\hat{a}w + \hat{a}^{\dagger}z)^n |0\rangle
=  
\sum_{m=0}^{\lfloor n/2 \rfloor}
\frac{(n!)^2}{(m!)^2 (n-2m)! \, 2^{2m}}
(\cosh r)^{2n-2m} (\sinh r)^{2m}.
\end{align}

Using the above equations, the expectation values of the relevant operators for independently squeezed light fields are given by
\begin{align}
\langle (\hat{a}_{S}^{\prime})^{\dagger}\hat{a}_{S}^{\prime} \rangle &= \sinh ^2(r_S) + 
2 n \beta 
\sum_{k=0}^{\lfloor m/2 \rfloor} \frac{(m!)^2 (\cosh{r_P})^{2k} (\sinh{r_P})^{2m-2k}}{(k!)^2(m-2k)!2^{2k}} 
\bigg[
n\sum_{j=0}^{\lfloor (n-1)/2 \rfloor} \frac{[(n-1)!]^2 (\cosh{r_S})^{2n-2j-2} (\sinh{r_S})^{2j}}{(j!)^2(n-2j-1)!2^{2j}}-  \notag
\\&
\sum_{j=0}^{\lfloor n/2 \rfloor} \frac{(n!)^2 (\cosh{r_S})^{2n-2j} (\sinh{r_S})^{2j}}{(j!)^2(n-2j)!2^{2j}}
\bigg] + 
\beta^2 n^2\sum_{t=0}^{n}\sum_{q=0}^{m}  \frac{(n!)^2 (m!)^2(-1)^t}{t!q![(n-t)!]^2[(m-q)!]^2} \times \notag
\\&
\sum_{k=0}^{\lfloor (2m-q)/2 \rfloor} \frac{[(2m-q)!]^2 (\cosh{r_P})^{2k} (\sinh{r_P})^{4m-2q-2k}}{(k!)^2(2m-q-2k)!2^{2k}} 
\sum_{j=0}^{\lfloor (2n-t-1)/2 \rfloor} \frac{[(2n-t-1)!]^2 (\cosh{r_S})^{4n-2t-2j-2} (\sinh{r_S})^{2j}}{(j!)^2(2n-t-2j-1)!2^{2j}}+ \notag
\\&
O(\beta^3)+O(\beta^4), \label{eq:SepNS}
\end{align}

\begin{align}
\langle (\hat{a}_{S}^{\prime})^{\dagger}\hat{a}_{S}^{\prime}(\hat{a}_{S}^{\prime})^{\dagger}\hat{a}_{S}^{\prime} \rangle &= \sinh ^4(r_S)+2 \sinh ^2(r_S) \cosh ^2(r_S) + 
\notag
\\
&
2 \beta 
\sum_{k=0}^{\lfloor m/2 \rfloor} \frac{(m!)^2 (\cosh{r_P})^{2k} (\sinh{r_P})^{2m-2k}}{(k!)^2(m-2k)!2^{2k}} 
\bigg[ 
(4n^2+2n) \sum_{j=0}^{\lfloor (n)/2 \rfloor} \frac{(n!)^2 (\cosh{r_S})^{2n-2j} (\sinh{r_S})^{2j}}{(j!)^2(n-2j)!2^{2j}} - \notag
\\
&
2 n \sum_{j=0}^{\lfloor (n+1)/2 \rfloor} \frac{[(n+1)!]^2 (\cosh{r_S})^{2n-2j+2} (\sinh{r_S})^{2j}}{(j!)^2(n-2j+1)!2^{2j}}-2 n^3 \sum_{j=0}^{\lfloor (n-1)/2 \rfloor} \frac{[(n-1)!]^2 (\cosh{r_S})^{2n-2j-2} (\sinh{r_S})^{2j}}{(j!)^2(n-2j-1)!2^{2j}} 
\bigg]+  \notag
\\
&
O(\beta^2)+O(\beta^3)+O(\beta^4),   \label{eq:SepNS2}
\end{align}

\begin{align}
\langle (\hat{a}_{P}^{\prime})^{\dagger}\hat{a}_{P}^{\prime} \rangle &= \sinh ^2(r_P) + 
2 m \beta 
\sum_{k=0}^{\lfloor m/2 \rfloor} \frac{(m!)^2 (\cosh{r_P})^{2k} (\sinh{r_P})^{2m-2k}}{(k!)^2(m-2k)!2^{2k}} 
\sum_{j=0}^{\lfloor n/2 \rfloor} \frac{(n!)^2 (\cosh{r_S})^{2n-2j} (\sinh{r_S})^{2j}}{(j!)^2(n-2j)!2^{2j}} + \notag
\\&
\beta^2 m^2
\sum_{t=0}^{n}\sum_{q=0}^{m-1} \frac{(n!)^2 [(m-1)!]^2(-1)^t}{t!q![(n-t)!]^2[(m-q-1)!]^2}
\sum_{k=0}^{\lfloor (2m-q-1)/2 \rfloor} \frac{[(2m-q-1)!]^2 (\cosh{r_P})^{4m-2q-2k-2} (\sinh{r_P})^{2k}}{(k!)^2(2m-q-2k-1)!2^{2k}} \times   \notag
\\&
\sum_{j=0}^{\lfloor (2n-t)/2 \rfloor} \frac{[(2n-t)!]^2 (\cosh{r_S})^{2j} (\sinh{r_S})^{4n-2t-2j}}{(j!)^2(2n-t-2j)!2^{2j}}
+O(\beta^3)+O(\beta^4),  \label{eq:SepNP}
\end{align}

\begin{align}
\langle (\hat{a}_{P}^{\prime})^{\dagger}\hat{a}_{P}^{\prime}(\hat{a}_{P}^{\prime})^{\dagger}\hat{a}_{P}^{\prime} \rangle &= \sinh^4(r_P)+2 \sinh ^2(r_P) \cosh ^2(r_P) + 
\notag
\\
& 
4 \beta m \bigg[ 
\sum_{k=0}^{\lfloor (m+1)/2 \rfloor} \frac{[(m+1)!]^2 (\cosh{r_P})^{2m-2k+2} (\sinh{r_P})^{2k}}{(k!)^2(m-2k+1)!2^{2k}}
+ 
m \sum_{k=0}^{\lfloor m/2 \rfloor} \frac{(m!)^2 (\cosh{r_P})^{2k} (\sinh{r_P})^{2m-2k}}{(k!)^2(m-2k)!2^{2k}} \bigg] \times \notag
\\&
\sum_{j=0}^{\lfloor n/2 \rfloor} \frac{(n!)^2 (\cosh{r_S})^{2n-2j} (\sinh{r_S})^{2j}}{(j!)^2(n-2j)!2^{2j}}+
O(\beta^2)+O(\beta^3)+O(\beta^4).  \label{eq:SepNP2}
\end{align}


\section{Calculated quantities for the case of jointly squeezing} \label{sec:observablesJoin}

It is not possible to directly use the relations derived in Appendix \ref{app:analyticalObs} to obtain closed-form expressions for the expectation values of different observables for jointly squeezed fields. This limitation arises due to modifications of the commutation relations at higher orders. Nevertheless, we obtain the following closed-form expressions up to first order in $\beta$:  

\begin{align}
\langle (\hat{a}_{S}^{\prime})^{\dagger}\hat{a}_{S}^{\prime} \rangle &=\sinh^2(r)+2 \beta  n \left((m+n)! \cosh ^{2 m}(r) \sinh ^{2 n}(r)-n (m+n-1)! \cosh ^{2 m}(r) \sinh ^{2 (n-1)}(r)\right)+O(\beta^2),
\end{align}

\begin{align}
\langle (\hat{a}_{P}^{\prime})^{\dagger}\hat{a}_{P}^{\prime} \rangle &=\sinh^2(r)-2 \beta  m (m+n)! \sinh ^{2 m}(r) \cosh ^{2 n}(r)+O(\beta^2),
\end{align}

\begin{align}
\langle (\hat{a}_{P}^{\prime})^{\dagger}\hat{a}_{P}^{\prime} (\hat{a}_{S}^{\prime})^{\dagger}\hat{a}_{S}^{\prime} \rangle &=\sinh ^4(r)+\sinh ^2(r) \cosh ^2(r)+2 \beta  n \bigg( \sinh ^{2 m}(r) \cosh ^{2 n}(r) \left(m^2 \cosh ^2(r) (m+n-1)!+C(n,m) \sinh ^2(r)\right)+
\notag
\\& 
n^2 (m+n-1)! \sinh ^{2 (m+2)}(r) \cosh ^{2 (n-1)}(r)\bigg)-
2 \beta  m(m+n)! \sinh ^{2 m}(r) \cosh ^{2 n}(r) \left(m \cosh ^2(r)+(n+1) \sinh ^2(r)\right)+O(\beta^2),
\end{align}   
in which $C(n,m)$ is a combinatorial factor arising from the normal ordering of the jointly squeezed moments, depending only on the photon orders $n$ and $m$.

\section{Scaling study} \label{sec:scaling_study}

To analyze the different limiting cases, we first recall the standard relations for the 
hyperbolic functions, $\sinh r = \frac{e^r - e^{-r}}{2}$ and $\cosh r = \frac{e^r + e^{-r}}{2}$. 
In the large-squeezing regime, $r \gg 1$, both functions grow exponentially, and we can 
approximate $\sinh r \approx \frac{e^r}{2}$ and $\cosh r \approx \frac{e^r}{2}$. This 
corresponds to the high-gain regime, in which multiphoton nonlinearities dominate and 
higher-order photon-number contributions become increasingly important.

In contrast, for small squeezing, $r \leqslant  1$, the functions can be expanded as 
$\sinh r \approx r + \frac{r^3}{6}$ and $\cosh r \approx 1 + \frac{r^2}{2}$, defining 
the low-gain regime where single- and few-photon processes prevail, and the system behaves 
approximately linearly. The intermediate range of $r$ between these two limits is treated 
as a mid-gain regime, in which both linear and nonlinear effects are relevant and compete 
to determine the photon statistics and correlations. We will not focus on this regime. 

These approximations provide a practical recipe for exploring the four limiting cases 
(low-low, low-high, high-low, high-high gain) introduced earlier, and for understanding 
how the nonlinear nature of the multiphoton absorption-emission process evolves across 
different squeezing strengths. The results for photon counting when the fields are independently squeezed are presented below 
\begin{align}
\Delta \beta_{N_S}\bigg|_{r_S\leqslant 1,r_P\leqslant 1}=\frac{2^{4 n - 5} r_S^2 (r_S^2 + 6)^2 (\frac {r_P^3} {6} + 
    r_P)^{-4 m} (r_S^2 + 2)^{6 - 4 n}}{n^2 (m!)^2 \Gamma (n)^2 \, _2F_1\left(\frac{1-m}{2},-\frac{m}{2};1;\frac {9 (r_P^2 + 2)^2} {r_P^2 (r_P^2 + 6)^2}\right)^2 \left(n (r_S+2)^2 \,
   _2F_1\left(\frac{1-n}{2},-\frac{n}{2};1;\frac {r_S^2 (r_S^2 + 6)^2} {9 (r_S^2 + 2)^2}\right)-4 n \, _2F_1\left(\frac{1-n}{2},1-\frac{n}{2};1;\frac {r_S^2 (r_S^2 + 6)^2} {9 (r_S^2 + 2)^2}\right)\right)^2 },
\end{align}

\begin{align}
\Delta \beta_{N_S}\bigg|_{r_S \gg 1,r_P\leqslant 1}=\frac {\pi  2^{2 n - 
      5} (r_P+\frac{r_P^6}{3})^{-4 m} e^{-4 (n - 
        1) r_S}} {n^2 (m!)^2 \Gamma (n + \frac {1} {2})^2 \,
     _ 2 F_ 1 (\frac {1 - m} {2}, -\frac {m} {2}; 
        1;\frac {9 (r_P^2 + 2)^2} {r_P^2 (r_P^2 + 6)^2})^2},
\end{align}

\begin{align}
\Delta \beta_{N_S}\bigg|_{r_S \leqslant 1,r_P \gg 1}=\frac{\pi 2^{2 m + 4 n - 5}  r_S^2 (r_S^2 + 6)^2 (r_S^2 + 
    2)^{6 - 4 n} e^{-4 m r_P}}{9 n^4\Gamma \left(m+\frac{1}{2}\right)^2 \Gamma (n)^2 \left(n (r_S+2)^2 \,
   _2F_1\left(\frac{1-n}{2},-\frac{n}{2};1;\frac {r_S^2 (r_S^2 + 6)^2} {9 (r_S^2 + 2)^2}\right)-4 n \, _2F_1\left(\frac{1-n}{2},1-\frac{n}{2};1;\frac {r_S^2 (r_S^2 + 6)^2} {9 (r_S^2 + 2)^2}\right)\right)^2},
\end{align}

\begin{align}
\Delta \beta_{N_S}\bigg|_{r_S \gg 1,r_P \gg 1}=\frac{\pi ^2 2^{2 m+2 n-5} e^{-4 (m r_P+(n-1) r_S)}}{n^2 \Gamma \left(m+\frac{1}{2}\right)^2 \Gamma \left(n+\frac{1}{2}\right)^2},
\end{align}
and similarly, for photon differences, we find
\begin{align}
\Delta \beta_{N_S-N_P}\bigg|_{r_S\leqslant 1,r_P\leqslant 1}=\frac{2^{4 n - 5} (\frac {r_P^3} {6} + 
    r_P)^{-4 m} (r_P^2 (r_P^4 + 8 r_P^2 + 12)^2 + 
   r_S^2 (r_S^4 + 8 r_S^2 + 12)^2)}{A_1},
\end{align}
in which 
\begin{align}
A_1&=n^2 (r_S+2)^{4 n-4} \Gamma (m+1)^2 \Gamma (n)^2 \, _2F_1\left(\frac{1-m}{2},-\frac{m}{2};1;\frac {9 (r_P^2 + 2)^2} {r_P^2 (r_P^2 + 6)^2}\right)^2
\notag
\\
&
   \left((m+n) (r_S+2)^2 \, _2F_1\left(\frac{1-n}{2},-\frac{n}{2};1;\frac {r_S^2 (r_S^2 + 6)^2} {9 (r_S^2 + 2)^2}\right)-4 n \, _2F_1\left(\frac{1-n}{2},1-\frac{n}{2};1;\frac {r_S^2 (r_S^2 + 6)^2} {9 (r_S^2 + 2)^2}\right)\right)^2,
\end{align}   

\begin{align}
\Delta \beta_{N_S-N_P}\bigg|_{r_S \gg 1,r_P\leqslant 1}=\frac {\pi  2^{2 n - 5} (\frac {r_P^3} {6} + 
      r_P)^{-4 m} e^{-4 n r_S} (r_P^2 (r_P^4 + 8 r_P^2 + 12)^2 + 
     9 e^{4 r_S})} {9 (m!)^2 (m + n)^2 \Gamma (n + \frac {1} {2})^2 \,
     _ 2 F_ 1 (\frac {1 - m} {2}, -\frac {m} {2}; 
       1;\frac {9 (r_P^2 + 2)^2} {r_P^2 (r_P^2 + 6)^2})^2},
\end{align}

\begin{align}
\Delta \beta_{N_S-N_P}\bigg|_{r_S \leqslant 1,r_P \gg 1}=\frac{\pi  2^{2 m+4 n-5} e^{-4 m r_P} (r_S+2)^{4-4 n} \left(9 e^{4 r_P} + r_S^2 (r_S^4 + 8 r_S^2 + 12)^2\right)}{\Gamma \left(m+\frac{1}{2}\right)^2 \Gamma (n+1)^2 \left| (m+n) (r_S+2)^2 \,
   _2F_1\left(\frac{1-n}{2},-\frac{n}{2};1;\frac {r_S^2 (r_S^2 + 6)^2} {9 (r_S^2 + 2)^2}\right)-4 n \, _2F_1\left(\frac{1-n}{2},1-\frac{n}{2};1;\frac {r_S^2 (r_S^2 + 6)^2} {9 (r_S^2 + 2)^2}\right)\right|^2},
\end{align}

\begin{align}
\Delta \beta_{N_S-N_P}\bigg|_{r_S \gg 1,r_P \gg 1}=\frac{\pi ^2 2^{2 m+2 n-5} \left(e^{4 r_P}+e^{4 r_S}\right) e^{-4 (m r_P+n r_S)}}{(m+n)^2 \Gamma \left(m+\frac{1}{2}\right)^2 \Gamma \left(n+\frac{1}{2}\right)^2},
\end{align}
and finally, for photon correlation, we find 
\begin{align}
\Delta \beta_{N_P N_S}\bigg|_{r_S\leqslant 1,r_P\leqslant 1}=\frac {81\ 
     4^{2 n + 1} (\frac {r_P^3} {6} + r_P)^{-4 m} (r_S^2 + 2)^{4 - 
       4 n} (\frac {r_P^2 (r_P^2 + 6)^2 (r_P^6 + 30 r_P^4 + 108 r_P^2 + 72) r_S^2 (r_S^2 + 6)^2 (r_S^6 + 30 r_S^4 + 108 r_S^2 + 72)} {1679616} - (\frac {r_P^3} {6} + r_P)^4 (\frac {r_S^3} {6} + r_S)^4)} {n^2 \Gamma (m + 1)^2 \Gamma (n)^2 A_2^2},
\end{align}
in which
\begin{align}
A_2&=-9 m (n + 1) (r_S^2 + 2)^2 \, _ 2 F_ 1 (\frac {1 - m} {2}, -\frac {m} {2}; 1;\frac {9(r_P^2 + 2)^2} {r_P^2 (r_P^2 + 6)^2}) \times
\notag
\\
&
\bigg( (r_S^2 + 2)^2 \, _ 2 F_ 1 (\frac {1} {2}(-n - 1), -\frac {n}{2}; 1;\frac {r_S^2 (r_S^2 + 6)^2} {9 (r_S^2 + 2)^2}) - 4 \, _ 2 F_ 1 (\frac {1 - n} {2}, -\frac {n}{2};1;\frac {r_S^2 (r_S^2 + 6)^2} {9(r_S^2 + 2)^2}) \bigg) + 
\notag
\\
&
\bigg((m + 1) r_P^2 (r_P^2 + 6)^2 \, _ 2 F_ 1 (\frac {1} {2} (-m - 1), -\frac {m} {2};1;\frac {9 (r_P^2 + 2)^2} {r_P^2 (r_P^2 + 6)^2}) + 36 m \,_ 2 F_ 1 (\frac {1 - m} {2}, -\frac {m} {2}; 1;\frac {9 (r_P^2 + 2)^2} {r_P^2 (r_P^2 + 6)^2}) \bigg) \times
\notag
\\
&
\bigg(4 n \, _ 2 F_ 1 (\frac {1 - n} {2}, 1 - \frac {n} {2}; 1;\frac {r_S^2 (r_S^2 + 6)^2} {9(r_S^2 + 2)^2}) - n (r_S^2 + 2)^2 \, _ 2 F_ 1 (\frac {1 - n} {2}, -\frac {n} {2}; 1;\frac {r_S^2 (r_S^2 + 6)^2} {9 (r_S^2 + 2)^2} \bigg), \notag
\end{align}

\begin{align}
\Delta \beta_{N_P N_S}\bigg|_{r_S \gg 1,r_P\leqslant 1}=\frac{\pi  2^{2 n-7} r_P^2 \left(r_P^2+3\right) \left(r_P^2+6\right)^2 \left(r_P^4+36 r_P^2+36\right)
   \left(\frac{r_P^3}{6}+r_P\right)^{-4 m} e^{-4 n r_S}}{81 m^4 \Gamma (m)^2 \Gamma \left(n+\frac{3}{2}\right)^2 \,
   _2F_1\left(\frac{1-m}{2},-\frac{m}{2};1;\frac{9 \left(r_P^2+2\right)^2}{r_P^2 \left(r_P^2+6\right)^2}\right)^2},
\end{align}

\begin{align}
\Delta \beta_{N_P N_S}\bigg|_{r_S \leqslant 1,r_P \gg 1}=\frac {e^{4 r_P} r_S^2 (r_S^2 + 3) (r_S^2 + 6)^2 (r_S^4 + 36 r_S^2 + 36) 2^{4 m + 4 n - 5} (r_S^2 + 2)^{4 - 4 n}} {81 A_3^2},
\end{align}
in which
\begin{align}
A_3&=m (r_S^2 + 2)^2 m! e^{2 m r_P} \,_ 2 F_ 1 (\frac {1 - m} {2}, -\frac {m} {2}; 1;\frac {(1 + e^{2 r_P})^2} {(-1 + e^{2 r_P})^2}) \bigg((r_S^2 + 2)^2 (n + 1)! \, _ 2 F_ 1 (\frac {1} {2} (-n - 1), -\frac {n} {2}; 1;\frac {r_S^2(r_S^2 + 6)^2} {9 (r_S^2 + 2)^2}) - 
\notag
\\
&
4 (n + 1) n! \, _ 2 F_ 1 (\frac {1 - n} {2}, -\frac {n} {2}; 1;\frac {r_S^2 (r_S^2 + 6)^2} {9 (r_S^2 + 2)^2}) \bigg) - 
n e^{2 (m - 1) r_P} \bigg((e^{2 r_P} - 1)^2 (m + 1)! \, _ 2 F_ 1 (\frac {1} {2} (-m - 1), -\frac {m} {2}; 1;\frac {(1 + e^{2 r_P})^2} {(-1 + e^{2 r_P})^2}) + 
\notag
\\
&
4 m e^{2 r_P} m! \,_ 2 F_ 1 (\frac {1 - m} {2}, -\frac {m} {2}; 1;\frac {(1 + e^{2 r_P})^2} {(-1 + e^{2 r_P})^2})\bigg) 
\bigg(4 n (n - 1)! \, _ 2 F_ 1 (\frac {1 - n} {2}, 1 - \frac {n} {2}; 1;\frac {r_S^2 (r_S^2 + 6)^2} {9(r_S^2 + 2)^2}) - (r_S^2 + 2)^2 n! \, _ 2 F_ 1 (\frac {1 - n} {2}, -\frac {n} {2}; 1;\frac {r_S^2 (r_S^2 + 6)^2} {9 (r_S^2 + 2)^2})\bigg),  \notag
\end{align}

\begin{align}
\Delta \beta_{N_P N_S}\bigg|_{r_S \gg 1,r_P \gg 1}=
\frac{\pi ^2 2^{2 m+2 n-3} e^{4 (-m r_P-n r_S+r_P+r_S)}}{\Gamma \left(m+\frac{1}{2}\right)^2 \Gamma
   \left(n+\frac{1}{2}\right)^2 \left((2 m+1) n e^{2 r_P}+m (2 n+1) e^{2 r_S}\right)^2}. 
   \label{eq:sepcor}
\end{align}

Similarly, for the jointly squeezed light fields case, we have the following scaling behaviors

\begin{align}
\Delta \beta_{N_S}\bigg|_{r \leqslant 1}=
\frac{9\ 4^{2 n-3} r^2 \left(r^2+6\right)^2 \left(\frac{r^3}{6}+r\right)^{-4 m} \left(r^2+2\right)^{6-4 n}}{n^2 ((m+n-1)!)^2 \left(9 m \left(r^2+2\right)^2+n r^2 \left(r^2+6\right)^2\right)^2},
\end{align}

\begin{align}
\Delta \beta_{N_S}\bigg|_{r \gg 1}=
\frac{2^{4 m+4 n-6} e^{-4 r (m+n-1)}}{n^2 (m+n)^2 ((m+n-1)!)^2},
\end{align}

\begin{align}
\Delta \beta_{N_S-N_P}\bigg|_{r \leqslant 1}=\mathcal{O}(\beta),
\end{align}

\begin{align}
\Delta \beta_{N_S-N_P}\bigg|_{r \gg 1}=\mathcal{O}(\beta),
\end{align}

\begin{align}
\Delta \beta_{N_P N_S}\bigg|_{r \leqslant 1}=
\frac{4^{2 n-5} r^2 \left(r^4+8 r^2+12\right)^2 \left(r^{12}+34 r^{10}+385 r^8+1816 r^6+3432 r^4+1728 r^2+144\right) \left(\frac{r^3}{6}+r\right)^{-4 m} \left(r^2+2\right)^{-4 n}}{81 \left(A_4\right)^2},
\end{align}
in which 
\begin{align}
A_4&=
n \left(\frac{1}{4} m^2 \left(r^2+2\right)^2 (m+n-1)!+\frac{n^2 \left(r^3+6 r\right)^4 (m+n-1)!}{324 \left(r^2+2\right)^2}+C(n,m)
   \left(\frac{r^3}{6}+r\right)^2\right)-\frac{1}{36} m (m+n)! \left(9 m \left(r^2+2\right)^2+(n+1) r^2 \left(r^2+6\right)^2\right),
\end{align}

\begin{align}
\Delta \beta_{N_P N_S}\bigg|_{r \gg 1}=
\frac{5\ 2^{4 (m+n-1)} e^{-4 r (m+n-1)}}{\left(n \left(m^2+n^2\right) (m+n-1)!-m (m+n+1) (m+n)!+C(n,m) n\right)^2}.
\end{align}

\end{widetext}

\end{document}